\documentclass[10pt,twocolumn]{article}
\usepackage{wrapfig}
\usepackage{graphicx}
\usepackage{amsmath,amssymb}

\title{\textbf{On equilibrium charge distribution above dielectric surface}}

\author{Dmytro M. Lytvynenko and Yuriy V. Slyusarenko
    \vspace{3mm}
    \\
\emph{\small Akhiezer Institute for Theoretical Physics, NSC KIPT,
                1~Akademicheskaya str., 61108 Kharkiv, Ukraine}}
\date{}

\begin{document}
\twocolumn[\maketitle \small{
           The problem of the equilibrium state of the charged many-particle system above dielectric surface is formulated.We consider the case of the presence of the external attractive pressing field and the case of its absence. The equilibrium distributions of charges and the electric field, which is generated by these charges in the system in the case of ideally plane dielectric surface, are obtained. The solution of electrostatic equations of the system under consideration in case of small spatial heterogeneities caused by the dielectric surface, is also obtained. These spatial inhomogeneities can be caused both by the inhomogeneities of the surface and by the inhomogeneous charge distribution upon it. In particular, the case of the .wavy. spatially periodic surface is considered taking into account the possible presence of the surface charges.
           \vspace{3mm}
    \vspace{4mm}
    \\}]

\section{Introduction}

The problems concerned with the research of the charges above
dielectric surface belong to classical electrodynamics and
electrostatics problems. A special interest to such problems
appeared due to the phenomenon of the Wigner crystallization. These
researches were initiated in 1934 by Wigner in his theoretical work
\cite{wigner}, per se. In this work, the possibility of the
existence of periodic structures in the systems with repulsive
forces between particles was demonstrated by the example of the
crystallization of the three-dimensional low-density gas of
electrons in the field generated by the spatial-homogeneous positive
charge. This field played exactly a role of the compensative factor
for the repulsive forces. The Landau-Silin Fermi-liquid theory also
enables to predict the existence of spatially-periodic state of
electrons in metals and to describe its structure (see in this case
Ref. \cite{pps}). The experimental improvement of Wigners'
prediction of the three-dimensional crystallic structures still does
not exist (see, e.g., Refs. \cite{monarkha,mb}). This is caused by
difficulties in the achievement of the experimental conditions for
the mentioned phenomenon, which is also called as ''Wigner
crystallization''.

However, as it is well known, different two-dimension periodic
electron structures above the surface of a fluid helium are
experimentally realized (the so-called ''Wigner crystals''). The
works \cite{cole,shykin,brown,grimes} may be referred as the first
publications containing theoretical and real experimental results of
different properties of the surface electrons. The large number of
works related to the theoretical and experimental research in this
area has appeared by now.

The theoretical papers that are devoted to the microscopic
description of the charge state above dielectric surface are usually
based on the conception of an isolated charge above dielectric
surface interacting with its electrostatic reflection in dielectric
(''levitate electron'', see, e.g., Refs.
\cite{monarkha,mb,edelman}). In this case, the quantum-mechanical
state ''charge - electrostatic reflection'' is described as the
hydrogen like one-dimensional state with the corresponding energy
structure. Very often the localization of such quantum-mechanical
object in the ground state is considered (see Refs
\cite{monarkha,mb,cole,shykin,edelman}) occuring at some distance
from the surface (first ''Bohr radius''). This, particularly in most
cases, allows not to take into account the influence of the surface
inhomogeneity on the single charge state. However, at the
description of many-particle charge system above dielectric surface
the mentioned approach inevitably faces some difficulties. For
example, such difficulty appears when the electron density above
dielectric surface does not allow to consider the charged particles
as isolated, i.e., it is necessary to take into account the
interparticle interaction.

The references, which are devoted to the two-dimensional Wigner
crystallization in the phenomenological approach, predominantly
consider the system that consists of a large number of charged
particles near the surface of the fluid dielectric as a
two-dimensional structure (see, e.g., Refs.\cite
{monarkha,mb,cole,shykin,brown,grimes}).

Basing on the premises, it becomes clear that the complete
description of charges above dielectric surface needs to take into
account their spatial distribution in vacuum. The possibility of
charges adsorption by the surface must also be taken into
consideration (in this case the surface inhomogeneities must play a
crucial role). The possibility of charge spatial distribution above
dielectric surface comes from the fact that a charged particle is
always attracted by a dielectric surface. Moreover, in the
experiments \cite {monarkha,mb} concerned with the registration of
two-dimensional Wigner crystallization an external electric field
attracts charges to the surface and affects on their spatial
distribution.

The present paper is devoted to the problem of equilibrium charge
distribution above dielectric surface as in the external pressing
electric field as in its absence. This problem is considered in the
case of ideally plane vacuum-dielectric boundary and in the case of
''wavy'' spatial-periodic surface with account of the possibility of
existence of the''sticked'' to the surface charges. In our opinion,
the formulated problem  is interesting as from purely academic side
as from the research side of the influence of volume charges located
closely to the fluid helium surface on the spatial-inhomogeneous
state of the charges adsorbed on the helium surface.

\section{Equations of electrostatics for many-charge system above dielectric surface}

Let us consider the equilibrium system of charged particles
(Fermi-particles) with the charge $Q$ per particle that is situated
in vacuum above dielectric surface with the permittivity
$\mathcal{\varepsilon}$. We describe below the surface profile by
function $\mathcal{\xi}(\boldsymbol{\rho})\equiv
\mathcal{\xi}(\emph{x},\emph{y})$, where $\boldsymbol{\rho}\equiv
\{\emph{x},\emph{y}\}$ is the radius-vector in the plane
$\emph{z}=0$ of Cartesian coordinates
$\{\emph{x},\emph{y},\emph{z}\}$. The vacuum - dielectric boundary
lies in the plane $\emph{z}=0$  and we consider it unbounded below.
All physical quantities considered in the area above dielectric,
i.e., at $\emph{z}>\xi(\boldsymbol{\rho})$ we mark by the index
''1'' and all quantities concerned with the dielectric
($\emph{z}<\xi(\boldsymbol{\rho})$) we mark by the index ''2''. Let
us assume that the external pressing electric field $E$ acts on
particles and is directed along the z-axis. We also assume the
existence of some potential barrier that forbids the charges to
penetrate inside the dielectric.

As it is mentioned above, the charged particles are always attracted
by the dielectric. Therefore, even in the absence of the external
pressing electric field there is a reason to believe that there are
conditions under what the stable equilibrium distribution along
$\emph{z}$-axis is developed. To avoid the questions on the
repulsion of likely charged particles along the plane
$\boldsymbol{\rho}$ we shall consider the system located in a vessel
with the walls at $\rho\rightarrow\infty$. These walls forbid the
charges to leave the system. Let us describe the equilibrium charge
distribution above the dielectric surface by the distribution
function $f(\textbf{p};z,\boldsymbol{\rho})$.

The electric field potential ${\varphi}_{\text{i}}$ in vacuum above
the dielectric surface must satisfy the Poisson's equation
\begin{equation}\label{eq1}                                  
\Delta{\varphi}_{\text{1}}(z,\boldsymbol{\rho})=-4{\pi}Qn(z,\boldsymbol{\rho}){\theta}(z-\xi(\boldsymbol{\rho})),
\end{equation}where $\Delta$ is the Laplace operator,
\begin{equation}\label{eq2}
\Delta\equiv\frac{\partial^{2}}{\partial{z}^{2}}+\Delta_{\boldsymbol{\rho}},~~
\Delta_{\boldsymbol{\rho}}\equiv\frac{\partial^{2}}{\partial{x}^{2}}+\frac{\partial^{2}}{\partial{y}^{2}},
\end{equation}
${\theta}(z-\xi(\boldsymbol{\rho}))$ is the Heaviside function. In
eq.~\eqref{eq1} the quantity $n(z,\boldsymbol{\rho})$ is the charge
density above the dielectric surface, which can be expressed in
terms of the distribution function
$f(\textbf{p};z,\boldsymbol{\rho})$ as
\begin{equation}\label{eq3}
n(z,\boldsymbol{\rho})=
\int{d}^{3}{p}f(\textbf{p};z,\boldsymbol{\rho}).
\end{equation}
As charges are considered as Fermi-particles, the distribution
function $f(\textbf{p};z,\boldsymbol{\rho})$ has the following form:

\begin{equation}\label{eq4}
\begin{gathered}
f({\bf{p}};z,\boldsymbol{\rho})
= \frac{g}{{(2\pi \hbar )^3
}}\times\hfill
\\
\quad\times\left\{ {\exp \beta \left[ {\frac{{p^2 }}{{2m}}
+ Q\varphi _1
(z,\boldsymbol{\rho}) - \mu } \right] + 1} \right\}^{ - 1} ,
\end{gathered}
\end{equation}
where $g=(2S_{Q}+1)$, $S_{Q}$ is the spin of the charged particle,
$\beta=1/T$, $T$ is the temperature in the energy units,  $m$ is the
charge mass and $\mu$ is the chemical potential of charges. We
emphasize that taking into account relations~\eqref{eq3} and
\eqref{eq4} the eq.~\eqref{eq1} is often called the Thomas-Fermi
equation.

The electric field potential $\varphi_{2}$ is the result of charge
absence in the dielectric.  In the assumption of the dielectric
homogeneity and isotropy it must satisfy the Laplace's equation
\begin{equation}\label{eq5}
\varepsilon\Delta\varphi_{2}(z,\boldsymbol{\rho})=0.
\end{equation}

If the system is placed in the external static homogeneous electric
field, the potentials $\varphi_{1}$ and $\varphi_{2}$ can be written
in the form
\begin{equation}\label{eq6}
\varphi_{1}=\varphi_{1}^{(i)}+\varphi_{1}^{(e)},~~~\varphi_{2}=\varphi_{2}^{(i)}+\varphi_{2}^{(e)},
\end{equation}
where $\varphi_{1}^{(i)}$, $\varphi_{2}^{(i)}$ are the potentials
induced by the system of charges in vacuum and in the dielectric,
respectively, $\varphi_{1}^{(e)}$ and  $\varphi_{2}^{(e)}$  are the
potentials of the external field in vacuum and dielectric. According
to eqs.~\eqref{eq1},~\eqref{eq5}, these fields satisfy the following
equations:
\begin{equation}\label{eq7}
\begin{gathered}
\Delta{\varphi}_{\text{1}}^{(i)}(z,\boldsymbol{\rho})=-4{\pi}Qn(z,\boldsymbol{\rho}),~~
\Delta\varphi_{\text{1}}^{(e)}=0,~~~\\
\Delta\varphi_{\text{2}}^{(i)}=0,~~~\Delta\varphi_{\text{2}}^{(e)}=0.
\end{gathered}
\end{equation}

Eqs.~\eqref{eq1},~\eqref{eq5} for the potentials must be expanded
with the boundary conditions on the vacuum-dielectric border (as
usual, these conditions can be obtained directly from
eqs.~\eqref{eq1},~\eqref{eq5}, see e.g.\cite{tumm}):
\begin{equation}\label{eq8}
\begin{gathered}
\varphi_{1}(z,\boldsymbol{\rho})|_{z=\xi(\boldsymbol{\rho})}=
\varphi_{2}(z,\boldsymbol{\rho})|_{z=\xi(\boldsymbol{\rho})},\\
n_{i}(\boldsymbol{\rho}) \{\varepsilon\nabla_{i}\varphi_{2}(z,\boldsymbol{\rho})-
\varepsilon\nabla_{i}\varphi_{1}(z,\boldsymbol{\rho})\}_{z=\xi(\boldsymbol{\rho})}
\\
=4\pi\sigma(\boldsymbol{\rho},\xi(\boldsymbol{\rho}')),
\end{gathered}
\end{equation}
where $\textbf{n}(\boldsymbol{\rho})$ is the unit vector of the
surface normal in the point $\boldsymbol{\rho}$,
$\sigma(\boldsymbol{\rho},\xi(\boldsymbol{\rho}'))$ is the surface
charge density in the point $\boldsymbol{\rho}$ (here we emphasize
the functional dependence of this value on the surface profile
$\xi(\boldsymbol{\rho}')$ ). The surface charge density
$\sigma(\boldsymbol{\rho},\xi(\boldsymbol{\rho}'))$ must satisfy the
following relation:
\begin{equation}\label{eq9}
\int{d}S_{\xi}\sigma(\boldsymbol{\rho},\xi(\boldsymbol{\rho}'))=QN_{\xi},
\end{equation}
where $N_{\xi}$ is the complete charge number on the dielectric
surface and ${d}S_{\xi}$ is the surface element with the profile
$\xi(\boldsymbol{\rho}')$:
\begin{equation}\label{eq10}
{d}S_{\xi}=d^{2}\rho\sqrt{1+(\partial\xi(\boldsymbol{\rho})/\partial\rho)^{2}}.
\end{equation}

The surface charge density can appear due to several reasons. For
example, these charges may be specially placed on the dielectric
surface and can stay there for arbitrary long time. At that time,
the surface charges and charges above the dielectric surface can
differ in sign. But in this case we need to consider the possibility
of the formation of bound states of the opposite charged particles.
Taking into account the presence of such bound states represents a
separate rather complicated problem. In the present paper such case
of surface charges is not considered. The case, in which some part
of charges condenses on the surface from the volume distribution,
stays for some period of time and then back to the volume, is
possible too. In this case, the equilibrium distribution of charges
above the surface that coexist with the ''sticked'' surface charges
for some period of time (the lifetime of the charge staying on the
surface) is possible. Next, we shall take into account only the
possibility of the presence of the surface charges that have the
same sign as the volume ones.

It is easy to see that in eq.~\eqref{eq8} the directional cosines of
the surface normal vector $\textbf{n}(\boldsymbol{\rho})$ in the
point $\boldsymbol{\rho}$ play the main role: $\cos\nu$ is cosine of
the angle between the normal and z-axis, $\cos\lambda$ is cosine of
the angle between the normal and x-axis and $\cos\mu$ is cosine of
the angle between the normal and y-axis. In the case when the
surface profile is given explicitly (in our case
$z=\xi(\boldsymbol{\rho})$), these cosines are determined by the
following relations:
\begin{equation}\label{eq11}
\begin{gathered}
\cos\nu=\frac{1}{\sqrt{1+(\partial\xi(\boldsymbol{\rho})/\partial\rho)^{2}}},\\
\cos\lambda=-\frac{\partial\xi(\boldsymbol{\rho})/\partial{x}}
{\sqrt{1+(\partial\xi(\boldsymbol{\rho})/\partial\rho)^{2}}},\\
\cos\mu=-\frac{\partial\xi(\boldsymbol{\rho})/\partial{y}}
{\sqrt{1+(\partial\xi(\boldsymbol{\rho})/\partial\rho)^{2}}}.
\end{gathered}
\end{equation}

The derived electrostatics equations~\eqref{eq1},~\eqref{eq5} with
the boundary conditions~\eqref{eq8},~\eqref{eq11} can be solved
analytically in a very low case count. Some of these cases are
considered below. Before solving eqs.~\eqref{eq1},~\eqref{eq5}, let
us consider the simplification of the boundary
conditions~\eqref{eq8} in the case, when the surface profile differs
a little from the plane one. In this case we essentially have the
effective boundary conditions. From eqs.~\eqref{eq10},~\eqref{eq11}
it is obvious that the surface slightly differs from the plane one,
when the surface profile slowly varies on coordinate, i.e., when the
following inequalities take place:
\begin{equation}\label{eq12}
\left| {\partial \xi (\boldsymbol{\rho})/\partial x} \right| \ll 1
,~~~ \left| {\partial \xi (\boldsymbol{\rho})/\partial y} \right|
\ll 1 .
\end{equation}
Let us also consider that the surface profile
$\xi(\boldsymbol{\rho})$ can be given as:
\begin{equation}\label{eq13}
\xi (\boldsymbol{\rho}) = \xi  + \tilde \xi (\boldsymbol{\rho}),~~~
\left| \xi \right| \gg \left| {\tilde \xi (\boldsymbol{\rho})}
\right|.
\end{equation}
It is easy to see that the inequality~\eqref{eq12} is provided in
this case by the conditions
\begin{equation}\label{eq14}
\left| {\partial \tilde \xi (\boldsymbol{\rho})/\partial x} \right|
\ll 1,~~~ \left| {\partial \tilde \xi (\boldsymbol{\rho})/\partial
y} \right| \ll 1.
\end{equation}
The directional cosines~\eqref{eq11} with accuracy up to the second
order over
$\partial\widetilde{\xi}(\boldsymbol{\rho})/\partial{\boldsymbol{\rho}}$
have the following form:
\begin{equation}\label{eq15}
\begin{gathered}
\cos\nu\approx1,\quad
\cos\lambda=-\partial\widetilde{\xi}(\boldsymbol{\rho})/\partial{x},\\
\cos\mu=-\partial\widetilde{\xi}(\boldsymbol{\rho})/\partial{y}.
\end{gathered}
\end{equation}

If the relations~\eqref{eq13}-\eqref{eq15} take place, we can expect
that the charge and the field distributions in the system slightly
differ from the distributions that take place in the case of the
plane dielectric surface. Then, the potentials
$\varphi_{1}(z,\boldsymbol{\rho})$ and
$\varphi_{2}(z,\boldsymbol{\rho})$ (see
eqs.~\eqref{eq1},~\eqref{eq5}) can be written as
\begin{equation}\label{eq16}
\begin{gathered}
\varphi_{1}(z,\boldsymbol{\rho})=\varphi_{1}(z)+\widetilde{\varphi}_{1}(z,\boldsymbol{\rho}),\\
\varphi_{2}(z,\boldsymbol{\rho})=\varphi_{2}(z)+\widetilde{\varphi}_{2}(z,\boldsymbol{\rho}),
\end{gathered}
\end{equation}
where $\varphi_{1}(z)$ and $\varphi_{2}(z)$ are the potentials of
some electric field above the dielectric and inside of it (but not
on the surface!) in the case of the plane surface. The small
distortions of the field above the dielectric and inside of it are
described by the potentials
$\widetilde{\varphi}_{1}(z,\boldsymbol{\rho})$ and
$\widetilde{\varphi}_{2}(z,\boldsymbol{\rho})$ due to the surface
inhomogeneity in the mentioned above sense. The meaning of the
introduced potentials $\varphi_{1}(z)$ and $\varphi_{2}(z)$, and
also $\widetilde{\varphi}_{1}(z,\boldsymbol{\rho})$ and
$\widetilde{\varphi}_{2}(z,\boldsymbol{\rho})$ becomes more clear
after the obtaining of the Poisson's equations and effective
boundary conditions for them.

According to the assumption of small field pertrubations provided by
the wave surface, the following inequalities take place:
\begin{equation}\label{eq17}
\left| {\varphi _1 (z)} \right| \gg \left| {\tilde \varphi _1
(z,\boldsymbol{\rho})} \right| ,~~ \left| {\varphi _2 (z)} \right|
\gg \left| {\tilde \varphi _2 (z,\boldsymbol{\rho})} \right| .
\end{equation}
Let us also consider that the distribution of charges that can be
condensed on surface slightly differs from the homogeneous one:
\begin{equation}\label{eq18}
\begin{gathered}
\sigma(\boldsymbol{\rho},\xi)=\sigma(\xi)+\widetilde{\sigma}(\boldsymbol{\rho},\xi)
+\frac{\partial\sigma(\xi)}{\partial\xi}\widetilde{\xi}(\boldsymbol{\rho}),\\
\left| {\sigma (\xi )} \right| \gg \left| {\tilde \sigma
(\boldsymbol{\rho};\xi )} \right|,~ \left| {\sigma (\xi )} \right|
\gg \left| {\frac{{\partial \sigma (\xi )}}{{\partial \xi }}\tilde
\xi (\boldsymbol{\rho})} \right| .
\end{gathered}
\end{equation}
In the expressions~\eqref{eq18} the quantity
$\widetilde{\sigma}(\boldsymbol{\rho},\xi)$ corresponds to the
impact of the weakly inhomogeneous charge distribution on the plane
dielectric surface with $z=\xi$ profile. The quantity
$\widetilde{\xi}(\boldsymbol{\rho})[\partial\sigma(\xi)/\partial\xi]$
in eq.~\eqref{eq18} describes the surface charge inhomogeneity
related to the weak irregularity of the surface itself.

In the expressions~\eqref{eq16},~\eqref{eq17} we consider that in
the case of uniformly charged surface, which is ideally plane and
infinitely extended with charge density $\sigma(\xi)$ both the field
and the charge distributions are homogeneous along
$\boldsymbol{\rho}$ plane. In other words, the spatial charge and
field distribution depends only on z coordinates.

From the relations~\eqref{eq13}-\eqref{eq18} it is easy to obtain
the effective boundary conditions for field potentials on
vacuum-dielectric boundary in the case, when the dielectric surface
weakly differs from the ideally plane. To this end, we must develop
the perturbation theory over small values
$\widetilde{\xi}(\boldsymbol{\rho})$,
$\widetilde{\sigma}(\boldsymbol{\rho},\xi)$ and
$\partial\widetilde{\xi}(\boldsymbol{\rho})/\partial\boldsymbol{\rho},$
which, according to the expressions~\eqref{eq12}-\eqref{eq18}, can
be given as follows:
\begin{equation}\label{eq19}
\begin{gathered}
\biggl\{\varphi_{1}(z)+\widetilde{\varphi}_{1}(z,\boldsymbol{\rho})
\biggr\}_{\xi+\widetilde{\xi}(\boldsymbol{\rho})}=
\biggl\{\varphi_{2}(z)+\widetilde{\varphi}_{2}(z,\boldsymbol{\rho})
\biggr\}_{\xi+\widetilde{\xi}(\boldsymbol{\rho})},\hfill
\\
\biggl\{\varepsilon\frac{\partial}{\partial{z}}[\varphi_{2}(z)+\widetilde{\varphi}_{2}(z,\boldsymbol{\rho})]-
\frac{\partial}{\partial{z}}[\varphi_{1}(z)+\widetilde{\varphi}_{1}(z,\boldsymbol{\rho})]
\biggr\}_{\xi+\widetilde{\xi}(\boldsymbol{\rho})}
\\
=4\pi\sigma(\boldsymbol{\rho};\xi+\widetilde{\xi}(\boldsymbol{\rho})).
\end{gathered}
\end{equation}
Making the necessary calculations up to the first order of the
perturbation theory from the first relation of eq.~\eqref{eq19} we
obtain

\begin{equation}\label{eq20}
\begin{gathered}
\varphi_{1}(z)|_{z=\xi}=\varphi_{2}(z)|_{z=\xi},\\
\biggl\{\widetilde{\varphi}_{1}(z,\boldsymbol{\rho})-\widetilde{\varphi}_{2}(z,\boldsymbol{\rho}),
\biggr\}_{z=\xi}=\hfill
\\
\quad=
\biggl\{\frac{\partial\varphi_{2}(z)}{\partial{z}}-\frac{\partial\varphi_{1}(z)}{\partial{z}}
\biggr\}_{z=\xi} \widetilde{\xi}(\boldsymbol{\rho}).
\end{gathered}
\end{equation}
The use of the perturbation theory up to the first order for the
second relation of eq.~\eqref{eq19} results in the following
equalities:
\begin{equation}\label{eq21}
\begin{gathered}
\biggl\{\varepsilon\frac{\partial\varphi_{2}(z)}{\partial{z}}-\frac{\partial\varphi_{1}(z)}{\partial{z}}
\biggr\}_{z=\xi}=4\pi\sigma(\xi),\hfill
\\
\biggl\{\varepsilon\frac{\partial^{2}\varphi_{2}(z)}{\partial{z^{2}}}-
\frac{\partial^{2}\varphi_{1}(z)}{\partial{z^{2}}}-
4\pi\frac{\partial\sigma(\xi)}{\partial\xi}
\biggr\}_{z=\xi}\widetilde{\xi}(\boldsymbol{\rho})-
\\-
4\pi\widetilde{\sigma}(\boldsymbol{\rho};\xi) =
\biggl\{\frac{\partial\widetilde{\varphi}_{1}(z,\boldsymbol{\rho})}{\partial{z}}
-\varepsilon\frac{\partial\widetilde{\varphi}_{2}(z,\boldsymbol{\rho})}{\partial{z}}
\biggr\}_{z=\xi}.
\end{gathered}
\end{equation}
Let us remind that in electrostatics the denotations like
$\{\partial^{2}\varphi_{1}(z)/\partial{z^{2}}\}_{z=\xi}$,
$\{\partial^{2}\varphi_{2}(z)/\partial{z^{2}}\}_{z=\xi}$ have the
meaning of limits
\begin{equation*}
\begin{gathered}
\{\partial^{2}\varphi_{1}(z)/\partial{z^{2}}\}_{z=\xi}=
\lim_{h\rightarrow0}
\{\partial^{2}\varphi_{1}(z)/\partial{z^{2}}\}_{z=\xi+h},\\
\{\partial^{2}\varphi_{2}(z)/\partial{z^{2}}\}_{z=\xi}=
\lim_{h\rightarrow0}
\{\partial^{2}\varphi_{2}(z)/\partial{z^{2}}\}_{z=\xi-h}.
\end{gathered}
\end{equation*}
Then, for the further simplification of the obtained effective
boundary conditions~\eqref{eq20},~\eqref{eq21}, according to
eqs.~\eqref{eq1},~\eqref{eq5},~\eqref{eq16} we can use the following
equations, which are satisfied by the potentials $\varphi_{1}(z)$,
$\varphi_{2}(z)$:
\begin{equation}\label{eq22}
\frac{\partial^{2}}{\partial^{2}}\varphi_{1}(z)=-4\pi{Q}n(z)\theta(z-\xi),~~
 \frac{\partial^{2}}{\partial^{2}}\varphi_{2}(z)=0,
\end{equation}
where
\begin{equation}\label{eq23}
\begin{gathered}
n(z)=\int{d}^{3}pf(\textbf{p},z),\\
 f(\textbf{p};z)=
\frac{g}{(2\pi\hbar)^{3}}  \biggl\{
{\exp\beta\left[\frac{p^{2}}{2m}+Q\varphi_{1}(z)-\mu\right]+1}
\biggr\}^{-1}.
\end{gathered}
\end{equation}
Taking into account eq.~\eqref{eq21}, the conditions~\eqref{eq20}
can be expressed in the following form (the first one of them
remains the same):
\begin{equation}\label{eq24}
\begin{gathered}
\biggl\{\varepsilon\frac{\partial\varphi_{2}(z)}{\partial{z}}-\frac{\partial\varphi_{1}(z)}{\partial{z}}
\biggr\}_{z=\xi}=4\pi\sigma(\xi),\hfill
\\
4\pi\biggl\{Qn(z)- \frac{\partial\sigma(\xi)}{\partial\xi}
\biggr\}_{z=\xi}\widetilde{\xi}(\boldsymbol{\rho})-
4\pi\widetilde{\sigma}(\boldsymbol{\rho};\xi) =
\\=
\biggl\{\frac{\partial\widetilde{\varphi}_{1}(z,\boldsymbol{\rho})}{\partial{z}}
-\varepsilon\frac{\partial\widetilde{\varphi}_{2}(z,\boldsymbol{\rho})}{\partial{z}}.
\biggr\}_{z=\xi}
\end{gathered}
\end{equation}
Thus, we obtain the effective boundary
conditions~\eqref{eq20},~\eqref{eq24} for the fields in the system
of charges above the dielectric surface with the surface profile
that slightly differs from plane
\begin{equation}\label{eq25}
\begin{gathered}
\frac{\partial^{2}\widetilde{\varphi}_{\text{1}}(z,\boldsymbol{\rho})}{\partial{z}^{2}}+
\Delta_{\boldsymbol{\rho}}\widetilde{\varphi}_{\text{1}}(z,\boldsymbol{\rho})
=4{\pi}Q^{2}\frac{\partial{n}(z)}{\partial\mu}
\widetilde{\varphi}_{\text{1}}(z,\boldsymbol{\rho}), \\
\frac{\partial^{2}\widetilde{\varphi}_{\text{2}}(z,\boldsymbol{\rho})}{\partial{z}^{2}}+
\Delta_{\boldsymbol{\rho}}\widetilde{\varphi}_{\text{2}}(z,\boldsymbol{\rho})=0.
\end{gathered}
\end{equation}

\section{System of charges above the ideally plane dielectric surface}

It is easy to see that the obtained
equations~\eqref{eq22},~\eqref{eq23} of electrostatics and the
effective boundary conditions~\eqref{eq20},~\eqref{eq24} for these
equations are much simpler than the initial electrostatic
equations~\eqref{eq1},~\eqref{eq5} and the boundary
conditions~\eqref{eq8}. Firstly, to solve the equations that
determine the charge and the field distribution above the
vacuum-dielectric boundary one needs consider the case of ideally
plane surface of this border that lies in $z=\xi$. Let us start
withe considering the case of the surface in the absence of the
charge $\sigma=0$. Then, the solution of~\eqref{eq22} must satisfy
the following boundary conditions:
\begin{equation}\label{eq26}
\begin{gathered}
\varphi_{1}(z)|_{z=\xi}=\varphi_{2}(z)\mid_{z=\xi},
\\
 \biggl\{
\varepsilon\frac{\partial\varphi_{2}(z)}{\partial{z}}-
\frac{\partial\varphi_{2}(z)}{\partial{z}} \biggr\}_{z=\xi}=0.
\end{gathered}
\end{equation}
To simplify the further calculations, let us write the first formula
from eq.~\eqref{eq22} in the following form:
\begin{equation}\label{eq27}
\frac{\partial\varphi_{1}^{2}(z)}{\partial{z}^{2}}
=-4\pi{Q}\nu\int\limits_{0}^{\infty}d\varepsilon\varepsilon^{1/2}\{\
\exp\beta(\varepsilon-\psi)+1\},
\end{equation}
where we denote
\begin{equation}\label{eq28}
\psi(z)\equiv\mu-Q\varphi_{1}(z),~~~\nu\equiv(2m)^{3/2}/2\pi^{2}\hbar^{3}.
\end{equation}
Here we also consider the spin of a charged particle equal to $1/2$,
$\psi$ is so-called electrochemical potential.

Multiplying eq.~\eqref{eq27} by the derivative
$(\partial\varphi_{1}(z)/\partial{z})$ and using the following
equality
\begin{equation*}
\biggl(\frac{\partial\varphi_{1}}{\partial{z}}\biggr)
\frac{1}{e^{\beta(\varepsilon-\psi)}+1}=
-\frac{1}{\beta{Q}}\frac{\partial}{\partial{z}} \ln
\biggl[e^{-\beta(\varepsilon-\psi)}+1\biggr],
\end{equation*}
after simple calculations we obtain the first-order differential
equation:
\begin{equation*}
\biggl(\frac{\partial\varphi_{1}}{\partial{z}}\biggr)^{2}=
\frac{16\pi}{3}\nu\int\limits_{0}^{\infty}d\varepsilon\varepsilon^{3/2}
 \{e^{\beta(\varepsilon-\psi)}+1\}^{-1}+C,
\end{equation*}
where $C$ is an arbitrary integration constant. Thus, the need of
the following equation solving arises:
\begin{equation}\label{eq29}
\frac{\partial\varphi_{1}}{\partial{z}}=\pm\biggl\{
\frac{16\pi}{3}\nu\int\limits_{0}^{\infty}d\varepsilon\varepsilon^{3/2}
 \{e^{\beta(\varepsilon-\psi)}+1\}^{-1}+C\biggr\}^{1/2}.
\end{equation}
The sign before the square root in eq.~\eqref{eq29} must be chosen
from the following consideration. The force acting on the charges at
$z>\xi$ presses these charges to the dielectric surface. Thus, in
the case of positive charges above the dielectric we choose the
positive sign, and in the case of negative charges we choose the
negative one. Let us consider below the distribution of negative
charges above the dielectric surface, $Q=-e$, $e>0$. Hence, the
potential $\varphi_{1}$ satisfies the relation:
\begin{equation}\label{eq30}
\frac{\partial\varphi_{1}}{\partial{z}}=-\biggl\{
\frac{16\pi}{3}\nu\int\limits_{0}^{\infty}d\varepsilon\varepsilon^{3/2}
 \{e^{\beta(\varepsilon-\psi)}+1\}^{-1}+C\biggr\}^{1/2}.
\end{equation}
Now we make the following denotations:
\begin{equation}\label{eq31}
\begin{gathered}
\varphi_{1}(z=0)\equiv\varphi_{0},~
\psi(z=0)\equiv\mu+e\varphi_{1}(z=0),
\\
E_0  \equiv  - \left(
{\frac{{\partial \varphi _1 (z)}}{{\partial z}}} \right)_{z = 0}
\end{gathered}
\end{equation}
Let us remind that we consider the case of the electric forces that
attract charges to the dielectric surface. Thus, at
$z\rightarrow\infty$ there is no charges, $f({\bf{p}};z)\mathop  \to
\limits_{z \to \infty } 0 $, or
\begin{equation}\label{eq32}
\left\{ {\exp \beta \left( {\varepsilon  - \psi } \right) + 1}
\right\}^{ - 1} \mathop  \to \limits_{z \to \infty } 0.
\end{equation}
The action of the electrostatic image force along z-axis must vanish
at $z \to \infty $:
\begin{equation*}
\frac{{\partial \varphi _1^{(i)} (z)}}{{\partial z}}\mathop  \to
\limits_{z \to \infty } 0.
\end{equation*}
As the result, it is essential to say that at $z \to \infty$ the
following relation takes place:
\begin{equation}\label{eq33}
- \frac{{\partial \varphi _1 (z)}}{{\partial z}}\mathop \to
\limits_{z \to \infty }  - \frac{{\partial \varphi _1^{(e)}
(z)}}{{\partial z}} \equiv E,
\end{equation}
where $E$ is the external field intensity that attracts charges to
the dielectric surface.

At $z=0$ from eq.~\eqref{eq30} one can get
\begin{equation*}
E_0^2  = \frac{{16\pi }}{3}\nu \int\limits_0^\infty  {d\varepsilon
\varepsilon ^{3/2} \left\{ {\exp \beta \left( {\varepsilon  - \psi
_0 } \right) + 1} \right\}^{ - 1} }  + C.
\end{equation*}
On the other hand, from the same equation and taking into account
eqs. (32), (33) at $z \to \infty $ we obtain:
\begin{equation}\label{eq34}
C = E^2.
\end{equation}
Comparing the last two expressions, we come to the relation between
the constants $\psi_{0}$, $E_{0}$ (see eq.~\eqref{eq31})  and the
external electromagnetic field $E$:
\begin{equation}\label{eq35}
E_0^2  - E^2  = \frac{{16\pi }}{3}\nu \int\limits_0^\infty
{d\varepsilon \varepsilon ^{3/2} \left\{ {\exp \beta \left(
{\varepsilon  - \psi _0 } \right) + 1} \right\}^{ - 1} }.
\end{equation}
Then, after integration of the first expression from
eq.~\eqref{eq23} over z within the limits from $\xi$ to
$\varepsilon$ and using eqs.~\eqref{eq32},~\eqref{eq34}, we get:
\begin{equation}\label{eq36}
E_0  - E = 4\pi en_s,\quad e > 0,
\end{equation}
where $n_s$ is the number of the volume charges per unit of the
plane dielectric surface:
\begin{equation}\label{eq37}
\begin{gathered}
n_s  = \int\limits_\xi ^\infty  {dzn(z)} ,
\\
\quad n(z) = \nu
\int\limits_0^\infty  {d\varepsilon \varepsilon ^{1/2} \left\{ {\exp
\beta \left( {\varepsilon  - \psi (z)} \right) + 1} \right\}^{ - 1}
}.
\end{gathered}
\end{equation}
Let us emphasize that for the equilibrium charge system above
dielectric the value of the number $n_s$ depends neither on the
coordinates, neither on the fields' distribution. It is determined
only by the entire number $N$ of the charges above dielectric. We
also
point out that this value  characterizes the additional 
field intensity that presses the charges to the dielectric surface.
Besides that, this field is generated by the charges themselves.

Thus, eqs.~\eqref{eq35},~\eqref{eq36} allow to express the unknown
quantities $\psi_0$ and $E_0$ (integration constants of
eq.~\eqref{eq27}) in terms of the external pressing electric field
$E$ and the number of charges above the unit item of the dielectric
surface $n_s$ (see eq.~\eqref{eq37}).

The second equation in~\eqref{eq38} can be solved trivially in
general case, because the electric field intensity in dielectric
does not depend on $z$. Using the boundary conditions~\eqref{eq26},
we can express the potential of electric field in dielectric in the
following form:
\begin{equation}\label{eq38}
\varphi _2  =  - \frac{{E_0 }}{\varepsilon }z + \varphi _0, \quad
E_2  = \frac{{E_0 }}{\varepsilon },
\end{equation}
where $E_2$ is the electric field intensity in the dielectric, $E_0$
can be expressed from eqs.~\eqref{eq35},~\eqref{eq36} and
$\varphi_0$ is the potential on the surface. As in the electrostatic
case, the field potential is determined accurate within a constant.
Therefore, we set the potential $\varphi_0$  equal to zero below. In
this case, the value of the electrochemical potential on the
dielectric surface coincides with the chemical one:
\begin{equation}\label{eq39}
\psi (z = 0) \equiv \psi _0  = \mu,
\quad
\varphi _0  = 0 .
\end{equation}
Taking into account eq.~\eqref{eq34}, the spatial distribution of
the potential (see eq.~\eqref{eq30}) can be written as follows:
\begin{equation}\label{eq40}
\begin{gathered}
\frac{{\partial \varphi _1 }}{{\partial z}} =  - \left\{
{\frac{{16\pi }}{3}\nu \int\limits_0^\infty  {d\varepsilon
\varepsilon ^{3/2} 
}}\times\right.\hfill
\\
\quad\times\biggl.{{
\left\{ {\exp \beta \left( {\varepsilon  - \psi }
\right) + 1} \right\}^{ - 1} }  + E^2 } \biggr\}^{1/2},
\\
\quad \psi
(z) \equiv \mu  + e\varphi _1 (z).
\end{gathered}
\end{equation}
It is easy to see that in general case the solution of this equation
can be found only in quadratures (see below). However, the gas of
charged Fermi-particles above the dielectric surface is
nondegenerate. Therefore, the solution of eq.~\eqref{eq40} can be
obtained analytically. Indeed, in the case of nondegenerate gas its
distribution function has the form that weakly differs from
Boltzmans' one
\begin{equation*}
\left\{ {\exp \beta \left( {\varepsilon  - \psi } \right) + 1}
\right\}^{ - 1}  \sim \exp \beta \left( {\psi  - \varepsilon }
\right).
\end{equation*}
Accordingly, the expression for the density distribution of a gas
along the $z$ coordinate (see eq.~\eqref{eq37}) becomes:
\begin{equation}\label{eq41}
 n(z) \approx \frac{{\sqrt \pi  }}{2}\nu \beta ^{ - 3/2} \exp (\beta \psi
 ).
\end{equation}
As the Fermi-particle gas is degenerate at low temperature and high
density ranges (see e.g. \cite{landau}) from eq.~\eqref{eq41} one
can get the gas nondegeneracy condition:
\begin{equation*}
\exp (\beta \psi ) \ll 1,
\end{equation*}
As the electrochemical potential depends on z, this condition is
obviously realized in the case when the following inequality takes
place:
\begin{equation}\label{eq42}
\exp (\beta \psi_0 ) \ll 1,
\end{equation}
where $\psi_0$ is the electrochemical potential on the dielectric
surface (see eqs.~\eqref{eq28},~\eqref{eq39} in this case). The last
statement takes place because of the assumption of the particle
absence at $z\to\infty$, see above. So, according to
eq.~\eqref{eq41},~\eqref{eq42}, the formula~\eqref{eq40} can be
expressed as:
\begin{equation}\label{eq43}
\frac{{\partial \psi }}{{\partial z}} =
- \left\{ {4\pi ^{3/2} e^2 \beta ^{ - 5/2}
\nu \exp (\beta \psi ) + e^2 E^2 } \right\}^{1/2}.
\end{equation}
This equation has the analytical solution:
\begin{equation}\label{eq44}
\begin{gathered}
\frac{{\sqrt \pi  }}{2}\nu \beta ^{ - 3/2} \exp \left( {\beta \psi
(z)} \right) = \beta \frac{{E^2 }}{{8\pi }}\frac{{4\chi
(z)}}{{\left( {1 - \chi (z)} \right)^2 }},\\
\quad
\psi (z) \equiv \mu  + e\varphi _1 (z),
\end{gathered}
\end{equation}
where the function $\chi (z) $ is defined by the relation:
\begin{equation}\label{eq45}
\begin{gathered}
\chi (z) \equiv \frac{{E_0  - E}}{{E_0  + E}}\exp \left\{ { - (z - \xi )/z_0 } \right\},
\quad\\
z_0  \equiv (\beta eE)^{ - 1},
\quad
\beta ^{ - 1}  = T.
\end{gathered}
\end{equation}
Let us emphasize that the multiplier before the exponent in
eq.~\eqref{eq45} according to eq.~\eqref{eq36} can be expressed in
terms of the intensity of the external electric field and the number
of charges $n_s$ in the ''column'' above the surface unit element:
\begin{equation*}
\frac{{E_0  - E}}{{E_0  + E}} = \frac{{2\pi en_s }}{{E + 2\pi en_s
}}
\end{equation*}
From the eqs.~\eqref{eq41}, \eqref{eq43} and \eqref{eq44} follows
that the charge density above the dielectric surface has the
distribution:
\begin{equation}\label{eq46}
n(z) = \beta \frac{{E^2 }}{{8\pi }}\frac{{4\chi (z)}}{{\left( {1 - \chi (z)} \right)^2 }},
\end{equation}
and the electric field intensity above the dielectric $ E_1 (z) $ is
expressed as
\begin{equation}\label{eq47}
E_1 (z) = E\frac{{1 + \chi (z)}}{{1 - \chi (z)}}.
\end{equation}
It is easy to see that at high values of $z$, $ z \gg z_0 $ (see
eq.~\eqref{eq45}), the charge distribution above the dielectric
surface is close to the Boltzman distribution and the electric field
density exponentially tends to the external pressing electric field
density. This fact confirms the above assumptions(see
eqs.~\eqref{eq32}, \eqref{eq33}).

The inequality~\eqref{eq42} that determines the nondegeneracy
condition of the charge gas can be written in terms of the obtained
solutions:
 \begin{equation}\label{eq48}
en_s \nu ^{ - 1} \beta ^{5/2} \left( {E + 2\pi en_s } \right) \ll 1.
\end{equation}
It is obvious that this inequality is not accomplished in the case
of low temperature range or high values of the external pressing
field.

Expressions~\eqref{eq44}-~\eqref{eq48} allow to make the limit
process at $ E \to 0 $. In the case of the absence of the external
pressing field, these solutions have the following form:
\begin{equation}\label{eq49}
\begin{gathered}
E_1 (z)\mathop  \to \limits_{E \to 0} E_0 \left\{ {1 + \frac{{z - \xi }}{{2z_0 }}} \right\}^{ - 1},
\quad\\
n(z)\mathop  \to \limits_{E \to 0} \beta \frac{{E_0^2 }}{{8\pi }}\left\{ {1 + \frac{{z - \xi }}{{2z_0 }}} \right\}^{ - 2},
\quad
E_2  = E_0 /\varepsilon,
\end{gathered}
\end{equation}
where
\begin{equation}\label{eq50}
\begin{gathered}
z_0  \equiv (\beta eE_0 )^{ - 1},
\quad
E_0  = 4\pi en_s.
\end{gathered}
\end{equation}
If to compare the expressions~\eqref{eq46}, \eqref{eq47} and
\eqref{eq49}, it is easy to see that in the case of the absence of
the external pressing field the exponential law of the electric
field and charge density above the dielectric surface changes to the
weaker power depedence. In this case, the inequality~\eqref{eq48}
can be written as:
\begin{equation}\label{eq51}
\begin{gathered}
 \left( {en_s } \right)^2 \nu ^{ - 1} \beta
^{5/2}  \ll 1.
\end{gathered}
\end{equation}
Note, that it takes place in the region of the relatively high
temperatures and low charge number in the volume above a surface
area unit, see eq.~\eqref{eq37}.

The obtained formulae~\eqref{eq38}, \eqref{eq44}-\eqref{eq51} are
the solution of the problem of the field and nondegenerate charged
gas distribution in charged particle system above the plane
dielectric surface as in the external pressing field as in its
absence. Let us emphasize that
 the dielectric permittivity does not appear in these expressions. The
reason is that the problem is homogeneous along the surface
coordinate $\rho$. In the case of inhomogeneity along $\rho$, the
solution of equations essentially depends on the sort of the
dielectric, i.e., on its permittivity $\varepsilon$. These
inhomogeneities may be caused by the inhomogeneities of the surface
itself or by inhomogeneity charge distribution on it (or the both
reasons simultaneously, see eqs.~\eqref{eq8}, \eqref{eq11},
\eqref{eq20}, \eqref{eq24}).

In the case of the degenerate gas, i.e., when the
condition~\eqref{eq48} or~\eqref{eq51} fails, the solution that is
obtained earlier is inapplicable. Let us make the following remark
relating to this fact. As it is mentioned earlier, the charge
density distribution decreases with the distance from the surface.
For this reason, in general case described by eq.~\eqref{eq40} the
gas can be degenerate in the area near the dielectric surface and
nondegenerate far from it. The typical distance from the surface
that separates these cases can be obtained using the following
considerations. As it is well known (see e. g. Ref.~\cite{landau}),
in low temperatures region the temperature expansions are widely
used for the calculus of the thermodynamical quantities
characterizing the gas. Applying such expansion to the integral over
the energy in eq.~\eqref{eq40}, we obtain:
\begin{equation}\label{eq52}
\begin{gathered}
 \int\limits_0^\infty  {d\varepsilon
\varepsilon ^{3/2} \left\{ {\exp \beta \left( {\varepsilon  - \psi }
\right) + 1} \right\}^{ - 1} } \approx
\\
 \approx \frac{2}{5}\psi ^{5/2}  +
\frac{{\pi ^2 }}{4}\beta ^{ - 2} \psi ^{1/2}  - \frac{{7\pi ^4
}}{{960}}\beta ^{ - 4} \psi ^{ - 3/2}  + ...
\end{gathered}
\end{equation}
From this expression it is easy to see that such expansion is
absolutely useless near the  point $z_1 $ obtained from the
condition
\begin{equation}\label{eq53}
\psi (z_1 ) = \mu  + e\varphi _1 (z_1 ) = 0.
\end{equation}
The solution of eq.~\eqref{eq40} obtained in quadratures is given by
\begin{equation}\label{eq54}
\begin{gathered}
z - \xi  =  - z_0 \int\limits_{\beta \psi _0 }^{\beta \psi } {d\zeta
} \biggl\{ {\frac{{16\pi }}{3}\nu \beta ^{ - 5/2} E^{ - 2}
\int\limits_0^\infty  {dyy^{3/2}\times
}}\biggr.\hfill
\\
\quad\times\biggl.{{
 \left\{ {\exp \left( {y - \zeta }
\right) + 1} \right\}^{ - 1} }  + 1} \biggr\}^{ - 1/2},
\end{gathered}
\end{equation}
where the distance $z_0 $ is determined by eq.~\eqref{eq45}. Taking
into account eqs.~\eqref{eq53}, \eqref{eq54}, the expression of the
border distance $z_1$ can be written as
\begin{equation}\label{eq55}
\begin{gathered}
z_1  = \xi  + z_0 \int\limits_0^{\beta \psi _0 } {d\zeta } \left\{
{\frac{{16\pi }}{3}\nu \beta ^{ - 5/2} E^{ - 2} \int\limits_0^\infty
{dyy^{3/2} \times
}}\right.\hfill
\\
\quad\times\biggl.{{
\left\{ {\exp \left( {y - \zeta } \right) + 1} \right\}^{
- 1} }  + 1} \biggr\}^{ - 1/2},
\end{gathered}
\end{equation}
where the electrochemical potential $ \psi _0 $ as the function of
temperature $ T = \beta ^{ - 1}$ and the external electric field is
obtained from the equation (see eqs.~\eqref{eq35}, \eqref{eq36},
\eqref{eq39})
\begin{equation}\label{eq56}
\begin{gathered}
\left( {4\pi en_s  + E} \right)^2  - E^2  =
\\
=\frac{{16\pi }}{3}\nu
\int\limits_0^\infty  {d\varepsilon \varepsilon ^{3/2} \left\{ {\exp
\beta \left( {\varepsilon  - \psi _0 } \right) + 1} \right\}^{ - 1}
}.
\end{gathered}
\end{equation}
As it is mentioned above, the potential $ \varphi _1 (z)$
 is defined accurate within an arbitrary
constant, which can be set equal to zero. Hence, eq.~\eqref{eq56} is
the expression defining the chemical potential $\mu$, see
eq.~\eqref{eq39}.

As expected, the typical distance $ z_1$ (see eq.~\eqref{eq55}) is
defined by the temperature, the external pressing field and the
number of charges above the dielectric surface area unit. Thus, the
charge gas is nondegenerate in the region $ z \gg z_1 $ and
degenerate at $ z \ll z_1 $. Let us point out that the
solutions~\eqref{eq45}-\eqref{eq50} are obtained assuming the charge
gas nondegeneracy in the entire area above the dielectric surface.
Therefore, in general case the mentioned expressions describe the
charge system only in the region $ z \gg z_1 $. The charge gas above
the dielectric surface can be degenerate even in the case of the
absence of the external pressing field. It is easy to see if to
analyze the expression~\eqref{eq55} at $ E \to 0$ with the account
of eq.~\eqref{eq45}.

In the case of the generate charge gas above dielectric surface,
eq.~\eqref{eq40} according to eq.~\eqref{eq52} can be written in
more simple form:
\begin{equation}\label{eq57}
\begin{gathered}
\frac{{\partial \varphi _1 }}{{\partial z}} = - \left\{
{\frac{{32\pi }}{{15}}\nu \psi ^{5/2}  + E^2 } \right\}^{1/2}, 
\\
\quad
\psi (z) \equiv \mu  + e\varphi _1 (z).
\end{gathered}
\end{equation}
However, in this case eq.~\eqref{eq57} can not be solved
analytically, and the numerical integration methods are needed.

Let us show now the influence of the plane dielectric surface
charges on the obtained results in the present section of this
paper. It is well known that infinitely thin homogeneously charged
plate with charge density $\sigma (\xi ) $ induces the homogeneous
field intensity $E_\sigma   = 2\pi \sigma (\xi ) $ in vacuum (in
this case the expression $\sigma (\xi ) $ shows that the surface
plane is described by the equation $ z = \xi $). This field has the
opposite direction in the opposite sides of the plane. In the case
of the charged plane dielectric surface the situation is absolutely
similar. E.g., the negatively charged dielectric surface induces the
field intensity $ E_{2\sigma }^{}  = 2\pi \left| {\sigma (\xi )}
\right|/\varepsilon $ in the dielectric and $E_{1\sigma }^{}  = -
2\pi \left| {\sigma (\xi )} \right| $ above the dielectric surface
(see eq.~\eqref{eq24}). As mentioned earlier in the present paper,
we consider only the cases of the same signs of charges as on the
dielectric surface, as in the volume above it (in our case we
consider the negative charges). In this case, the field induced by
the surface charges repulses the volume charges from the surface.
So, the results obtained in the present partition remain useful if
we substitute the external electric field in vacuum $E$ for $ E -
2\pi \left| {\sigma (\xi )} \right| $ in the
expressions~\eqref{eq31}-\eqref{eq56},
\begin{equation}\label{eq58}
 E \to E - 2\pi \left| {\sigma (\xi )} \right|.
\end{equation}
It is easy to see that it is necessary to satisfy the condition
\begin{equation}\label{eq59}
E_0  - 2\pi \left| {\sigma (\xi )} \right| > 0
\end{equation}
that provides the possibility of existence of the equilibrium volume
charge distribution above the dielectric surface in the repulsive
field of the surface charges.

\section{The charge system above the spatially inhomogeneous dielectric surface}

As already mentioned, the spatial inhomogeneities can be caused by
the surface heterogeneities or by the inhomogeneous charge
distribution on it (or the both reasons simultaneously, see
eqs.~\eqref{eq8}, \eqref{eq11}, \eqref{eq20}, \eqref{eq24})). Let us
consider the mentioned surface inhomogeneities that slightly distort
the electric field induced by the charge system above the plane
dielectric:
\begin{equation*}
\begin{gathered}
\varphi _1 (z,\boldsymbol{\rho}) = \varphi _1 (z) + \tilde \varphi
_1 (z,\boldsymbol{\rho}),\quad\left| {\varphi _1 (z)} \right| \gg \left| {\tilde \varphi _1
(z,\boldsymbol{\rho})} \right|,
\\
\quad \varphi _2 (z,\boldsymbol{\rho}) =
\varphi _2 (z) + \tilde \varphi _2 (z,\boldsymbol{\rho}), \quad \left| {\varphi _2 (z)}
\right| \gg \left| {\tilde \varphi _2 (z,\boldsymbol{\rho})}
\right|,
\end{gathered}
\end{equation*}
where $ \varphi _1 (z) $, $ \varphi _2 (z) $ are the potentials
above the dielectric and inside of it, respectively, in the case of
the ideally plane dielectric surface with the equation of the
profile $ z = \xi $ (see eqs.~\eqref{eq16}-~\eqref{eq18}). The
obtaining procedure for the potentials $ \varphi _1 (z) $, $ \varphi
_2 (z) $ and charge density $ n(z) $ is described in details in the
previous section of the present paper, see
eqs.~\eqref{eq26}-~\eqref{eq59}.

The next problem is concerned with the potentials $ \tilde \varphi
_1 (z,\boldsymbol{\rho}) $ and $ \tilde \varphi _2
(z,\boldsymbol{\rho}) $ obtaining. For these potentials one can use
the eq.~\eqref{eq25} and the boundary conditions~\eqref{eq20},
\eqref{eq24}. In terms of the Fourier-transforms $\tilde \varphi _1
(z,{\bf{q}}) $ , $\tilde \varphi _2 (z,{\bf{q}}) $  over coordinate
$\boldsymbol{\rho}$ of the potentials $ \tilde \varphi _1
(z,\boldsymbol{\rho}) $ and $ \tilde \varphi _2
(z,\boldsymbol{\rho}) $
\begin{equation}\label{eq60}
\begin{gathered}
\tilde \varphi _1 (z,\boldsymbol{\rho}) = \int {d^2 q\exp \left( {i{\bf{q}\boldsymbol{\rho}}} \right)\tilde \varphi _1 (z,{\bf{q}})},
\\
\quad
\tilde \varphi _2 (z,\boldsymbol{\rho}) = \int {d^2 q\exp \left( {i{\bf{q\boldsymbol{\rho} }}} \right)\tilde \varphi _2 (z,{\bf{q}})}
\end{gathered}
\end{equation}
the equations~\eqref{eq25} have the following form:
\begin{equation}\label{eq61}
\begin{gathered}
\frac{{\partial ^2 \tilde \varphi _1 (z,{\bf{q}})}}{{\partial z^2 }}
- q^2 \tilde \varphi _1 (z,{\bf{q}}) = 4\pi e^2 \frac{{\partial
n(z)}}{{\partial \mu }}\tilde \varphi _1 (z,{\bf{q}}),
\\
\frac{{\partial ^2 \tilde \varphi _2 (z,{\bf{q}})}}{{\partial z^2 }}
- q^2 \tilde \varphi _2 (z,{\bf{q}}) = 0.
\end{gathered}
\end{equation}
According to eqs.~\eqref{eq20}, \eqref{eq24}, the boundary
conditions concerned with these equations can be written as:
\begin{equation}\label{eq62}
\begin{gathered}
\left\{ {\tilde \varphi _1 (z,{\bf{q}}) - \tilde \varphi _2
(z,{\bf{q}})} \right\}_{z = \xi } = \hfill
\\
= \quad\left\{ {\frac{{\partial
\varphi _2 (z)}}{{\partial z}} - \frac{{\partial \varphi _1
(z)}}{{\partial z}}} \right\}_{z = \xi } \tilde \xi ({\bf{q}}),
\\
- 4\pi \left\{ {en(z) + \frac{{\partial \sigma (\xi )}}{{\partial
\xi }}} \right\}_{z = \xi } \tilde \xi ({\bf{q}}) - 4\pi \tilde
\sigma ({\bf{q}};\xi )=\hfill 
\\
= \quad\left\{ {\frac{{\partial \tilde \varphi _1
(z,{\bf{q}})}}{{\partial z}} - \varepsilon \frac{{\partial \tilde
\varphi _2 (z,{\bf{q}})}}{{\partial z}}} \right\}_{z = \xi},
\end{gathered}
\end{equation}
where $ \tilde \xi ({\bf{q}}) $, $ \tilde \sigma ({\bf{q}};\xi ) $
are the Fourier-transforms of the quantities $ \tilde \xi
(\boldsymbol{\rho}) $ and $ \tilde \sigma (\boldsymbol{\rho};\xi )
$, respectively (see eqs. (14), (18)):
\begin{equation}\label{eq63}
\begin{gathered}
\tilde \xi (\boldsymbol{\rho}) = \int {d^2 q\exp \left(
{i{\bf{q\boldsymbol{\rho} }}} \right)\tilde \xi ({\bf{q}})}, 
\\
\tilde \sigma (\boldsymbol{\rho};\xi ) = \int {d^2 q\exp \left(
{i{\bf{q}\boldsymbol{\rho}}} \right)\tilde \sigma ({\bf{q}};\xi )}.
\end{gathered}
\end{equation}
Let us consider the electric fields intensity perturbations caused
by inhomogeneities of the dielectric surface rapidly decreasing at $
z \to \pm\infty $. It is easy to see that the first equation in
eq.~\eqref{eq61} in general case cannot be solved analytically. But
in two particular cases the analytical solution exists. In the first
case, we solve eq.~\eqref{eq61} at $ z \sim \xi$ setting $\partial
n(z)/\partial \mu$ equal to its value on the plane surface, $z = \xi
$:
\begin{equation}\label{eq64}
\frac{{\partial n(z)}}{{\partial \mu }} \approx \frac{{\partial
n(\xi )}}{{\partial \mu }}.
\end{equation}
Such consideration is possible in the case when the typical size of
spatial inhomogeneities of the unperturbed charge density $n(z) $
considerably larger than the typical size of the spatial
inhomogeneities of the potential $ \tilde \varphi _1 (z,{\bf{q}})$
along $z$-axis:
\begin{equation}\label{eq65}
\begin{gathered}
\left| {\left( {\frac{{\partial n(z)}}{{\partial \mu }}} \right)^{ -
1} \frac{\partial }{{\partial z}}\frac{{\partial n(z)}}{{\partial
\mu }}} \right|_{z = \xi }\ll \hfill
\\
\quad\ll \left| {\left\{ {\tilde \varphi _1
(z,{\bf{q}})} \right\}^{ - 1} \frac{{\partial \tilde \varphi _1
(z,{\bf{q}})}} {{\partial z}}} \right|_{z = \xi }.
\end{gathered}
\end{equation}
Let us return to the discussion of the condition~\eqref{eq64} below.

Then, taking into account the assumption of rapidly fading field
densities at $ z \to\pm\infty $, the solution of eq.~\eqref{eq60}
can be given in the following form:
\begin{equation}\label{eq66}
\begin{gathered}
\tilde \varphi _1 (z,{\bf{q}}) = A_1 ({\bf{q}})\exp \left( { -
zb(q)} \right), 
\\
\quad \tilde \varphi _2 (z,{\bf{q}}) = A_2
({\bf{q}})\exp \left( {zq} \right),
\end{gathered}
\end{equation}
where (see eq.~\eqref{eq64})
\begin{equation}\label{eq67}
b(q) \equiv \sqrt {q^2  + 4\pi e^2 \frac{{\partial n(\xi
)}}{{\partial \mu }}},
\end{equation}
and $ A_1 ({\bf{q}}) $, $ A_2 ({\bf{q}})$  are obtained from the
boundary conditions~\eqref{eq62}. To this end, we put the
expressions~\eqref{eq66} into the boundary conditions~\eqref{eq62}
and obtain the following relations for the potentials $ \tilde
\varphi _1 (z,{\bf{q}})$, $\tilde \varphi _2 (z,{\bf{q}}) $:
\begin{equation}\label{eq68}
\begin{gathered}
\tilde \varphi _1 (z,{\bf{q}})
= \frac{{\exp \left( { - (z - \xi )b(q)} \right)}}{{\varepsilon q + b(q)}}
\left\{ {\left[ {\varepsilon q\left( {E_1 (\xi ) - E_2 (\xi )} \right)} \right.}\right.
+\hfill
\\
\quad\left.{\left. { + 4\pi \left( {en(\xi ) +
\left( {\partial \sigma (\xi )/\partial \xi } \right)} \right)} \right]
\tilde \xi ({\bf{q}};\xi )+ 4\pi \tilde \sigma ({\bf{q}};\xi )} \right\} \hfill
,\\
\quad
\tilde \varphi _2 (z,{\bf{q}}) =
- \frac{{\exp \left( {(z - \xi )q} \right)}}{{\varepsilon q +
b(q)}}\left\{ {\left[ {b(q)\left( {E_1 (\xi ) - E_2 (\xi )} \right)}
\right.}\right.-\hfill
\\
\quad\left.
 {\left. { - 4\pi \left( {en(\xi ) + \left(
{\partial \sigma (\xi )/\partial \xi } \right)} \right)}
\right]\tilde \xi ({\bf{q}};\xi ) - 4\pi \tilde \sigma ({\bf{q}};\xi
)} \right\},
\end{gathered}
\end{equation}
where (see eqs.~\eqref{eq22}, \eqref{eq26})
\begin{equation*}
E_1 (\xi ) \equiv  - \left( {\frac{{\partial \varphi _1 (z)}}{{\partial z}}} \right)_{z = \xi }
,\quad
E_2 (\xi ) \equiv  - \left( {\frac{{\partial \varphi _2 (z)}}{{\partial z}}} \right)_{z = \xi }.
\end{equation*}
Let us emphasize that according to the boundary
conditions~\eqref{eq20}, \eqref{eq21} (see also eqs.~\eqref{eq58},
\eqref{eq59}) the values of the quantities $ E_1 (\xi ) $, $E_2 (\xi
)$ can be expressed as
\begin{equation}\label{eq69}
\begin{gathered}
E_1 (\xi ) = E_0  - 2\pi \left| {\sigma (\xi )} \right| > 0 ,\\
\quad
E_2 (\xi ) = \left( {E_0 + 2\pi \left| {\sigma (\xi )} \right|}
\right)/\varepsilon,
\end{gathered}
\end{equation}
where the field intensity $E_0$ is defined by the
relation~\eqref{eq36}:
\begin{equation*}
E_0  = E + 4\pi en_s.
\end{equation*}

Let us remind that the values of the potentials $ \tilde \varphi _1
(z,{\bf{q}}) $, $ \tilde \varphi _2 (z,{\bf{q}}) $ at $ z = \xi $ do
not coincide due to the fact that the potential continuity on the
surface in the case of its inhomogeneous wavy structure is provided
by the inequalities (see eqs.~\eqref{eq16}-\eqref{eq20}):
\begin{equation*}
\varphi _1 (z)\left| {_{z = \xi } } \right. = \varphi _2 (z)\left|
{_{z = \xi } } \right. ,\quad \delta \varphi _1 (\xi ,{\bf{q}}) =
\delta \varphi _2 (\xi ,{\bf{q}}),
\end{equation*}
where $ \delta \varphi _1 (\xi ,{\bf{q}}) \equiv \tilde \xi
({\bf{q}})\left( {\partial \varphi _1 (z)/\partial z} \right)_{z =
\xi }  + \tilde \varphi _1 (\xi ,{\bf{q}}) $. According to
eq.~\eqref{eq68}, one can get:
\begin{equation}\label{eq70}
\begin{gathered}
\delta \varphi _1 (\xi ,{\bf{q}}) = - \frac{1}{{\varepsilon q +
b(q)}}\biggl\{ {\left[ {\varepsilon qE_2 (\xi ) + b(q)E_1 (\xi ) - }
\right.} \biggr.\hfill
\\
\biggr.
\quad{\left. -{4\pi \left( {en(\xi ) + \left( {\partial \sigma
(\xi )/\partial \xi } \right)} \right)} \right]\tilde \xi
({\bf{q}};\xi ) - 4\pi \tilde \sigma ({\bf{q}};\xi )} \biggr\},
\end{gathered}
\end{equation}
where $ E_1 (\xi ) $, $ E_2 (\xi ) $ are still defined by the
relations (69). It is easy to see from the obtained
formulae~\eqref{eq68}, \eqref{eq70} that the gas of the volume
charges can sufficiently affect on the potential of the electric
field near the dielectric surface.

Now let us show that the solution of eq.~\eqref{eq61} in the
forms~\eqref{eq66}, \eqref{eq68} is correct. As it is mentioned
above, the condition of the existence of such solution is defined by
the relation~\eqref{eq65}. According to eq.~\eqref{eq67} it can be
expressed as follows:
 \begin{equation}\label{eq71}
\left| {\left( {\frac{{\partial n(\xi )}}{{\partial \mu }}}
\right)^{ - 1} \frac{\partial }{{\partial z}}\frac{{\partial n(\xi
)}}{{\partial \mu }}}\right| \ll \sqrt {q^2  + 4\pi e^2
\frac{{\partial n(\xi )}}{{\partial \mu }}}.
\end{equation}

The explicit expression for the derivative $\partial n(\xi
)/\partial \mu$ can be obtained from eqs.~\eqref{eq31},
\eqref{eq35}, \eqref{eq36}, \eqref{eq41}, \eqref{eq46},
\eqref{eq52}, \eqref{eq56} as in the case of degenerate charge gas
above the dielectric surface, as in the case of nondegenerate one.
In the second case the condition~\eqref{eq71} has a rather simple
form:
\begin{equation}\label{eq72}
q^2  \gg \beta ^2 e^2 \left\{ {\left( {E + 2\pi en_s } \right)^2  +
4\pi ^2 e^2 n_s^2 } \right\}.
\end{equation}
In the case, when the gas of charged Fermi-particles is degenerate
at $ z \ll z_1 $ (see eq.~\eqref{eq55}) and low temperature
expansions~\eqref{eq52} take place, we can obtain the following
expressions for the volume charge density $ n(z) $ at $ z \sim \xi $
and the electrochemical potential $ \psi _0 $ at $z = \xi $ (see
eqs.~\eqref{eq35}, \eqref{eq52}, \eqref{eq56}):
\begin{equation}\label{eq73}
n(z) \approx \frac{2}{3}\nu \psi ^{3/2},\quad
\psi _0  \approx \left\{ {\frac{{15}}{{32}}\frac{{E_0^2  - E^2 }}{{\pi \nu }}} \right\}^{2/5},
\end{equation}
where $\nu$ and $\psi$ are still defined by the
relations~\eqref{eq28} with $ Q = - e $, and $E_0$, $E$ are
expressed by~\eqref{eq36}. By the use of the
expression~\eqref{eq73}, one can write the condition~\eqref{eq71}:
\begin{equation}\label{eq74}
\begin{gathered}
E^2  - 52\pi
en_s E - 8\pi ^2 e^2 n_s^2\ll  \hfill
\\
\quad\ll 4q^2 e^{ - 2} \left\{
{\frac{{15}}{4}\frac{{en_s \left( {E + 2\pi en_s } \right)}}{\nu }}
\right\}^{4/5}.
\end{gathered}
\end{equation}

Accounting the solutions~\eqref{eq68} are the Fourier-transforms of
the potentials $ \tilde \varphi _1 (z,{\boldsymbol\rho}) $, $ \tilde
\varphi _2 (z,\boldsymbol\rho) $ (see eq.~\eqref{eq60}), the
relations~\eqref{eq71}-\eqref{eq74} in general case are correct for
any value of $q$, including also the value $q=0$. It is easy to see
that the relation~\eqref{eq72} does not satisfy such a requirement.
The relation~\eqref{eq74} can take place at all values of $q$ in the
case of the external pressing field $E$ that satisfies the following
inequality:
\begin{equation}\label{eq75}
0 \le E \le E',\quad E' \approx 52\pi en_s.
\end{equation}
At $ E > E' $ the expressions~\eqref{eq68} do not take place. In
this case, as in the case of the condition~\eqref{eq72} realization,
the equations~\eqref{eq61} must be solved by the use of the
numerical methods.

The case of the particular interest is the spatially periodic
inhomogeneities caused by the dielectric surface. As it is already
mentioned in the present paper, such inhomogeneities are concerned
with two-dimensional Wigner crystallization. In the most simple case
of spatial periodic inhomogeneities the Fourier-transforms of the
quantities $ \tilde \xi (\boldsymbol\rho ) $, $ \tilde \sigma
(\boldsymbol\rho ;\xi ) $ (see eq.~\eqref{eq63}) can be expressed in
the form:
\begin{equation}\label{eq76}
\begin{gathered}
\tilde \sigma ({\bf{q}};\xi ) = \frac{1}{2}\sum\limits_{\alpha  = 1}^2
{\tilde \sigma ({\bf{q}}_{\alpha \sigma } ;\xi )
\left\{ {\delta \left( {{\bf{q}} + {\bf{q}}_{\alpha \sigma } } \right)
+ \delta \left( {{\bf{q}} - {\bf{q}}_{\alpha \sigma } } \right)} \right\}},\\
\tilde \xi ({\bf{q}}) = \frac{1}{2}\sum\limits_{\alpha  = 1}^2
{\tilde \xi ({\bf{q}}_{\alpha \xi } )\left\{ {\delta \left( {{\bf{q}} + {\bf{q}}_{\alpha \xi } }
\right) + \delta \left( {{\bf{q}} - {\bf{q}}_{\alpha \xi } } \right)} \right\}},
\end{gathered}
\end{equation}
where ${\bf{q}}_{\alpha \sigma }$ $(\alpha  = 1,2) $ are the vectors
of the reciprocal two-dimensional lattice concerned with the spatial
periodic charge distribution on the dielectric surface,
${\bf{q}}_{\alpha \xi }$ $(\alpha  = 1,2) $ are the vectors of the
reciprocal two-dimensional lattice concerned with the spatial
periodic wavy surface type, and $ \tilde \sigma ({\bf{q}}_{\alpha
\sigma } ;\xi ) $ , $ \tilde \xi ({\bf{q}}_{\alpha \xi } ) $ are the
amplitudes of the corresponding surface heterogeneities. Of course,
it is necessary to consider that the conditions~\eqref{eq14} take
place, which in this case can be written as:
\begin{equation}\label{eq77}
q_{\alpha \xi } \tilde \xi ({\bf{q}}_{\alpha \xi } ) \ll 1, \quad
q_{\alpha \sigma } \tilde \xi ({\bf{q}}_{\alpha \sigma } ) \ll 1.
\end{equation}

Then, putting the expressions~\eqref{eq76} into eq.~\eqref{eq68} and
making inverse Fourier transformation according to eq.~\eqref{eq60},
it is easy to obtain the following expressions for the potentials $
\tilde \varphi _1 (z,\boldsymbol\rho) $, $ \tilde \varphi _2
(z,\boldsymbol\rho) $:
\begin{equation}\label{eq78}
\begin{gathered}
\tilde \varphi _1 (z,\boldsymbol\rho) = \sum\limits_{\alpha  = 1}^2
{\frac{{\exp \left( { - (z - \xi )b(q_{\alpha \xi } )} \right)}}
{{\varepsilon q_{\alpha \xi }  + b(q_{\alpha \xi } )}}} \biggl\{
{\varepsilon q_{\alpha \xi } \left(E_1 (\xi )-\right.} \biggr.\hfill
\\
\left.-E_2 (\xi )\right)+ \biggl. {4\pi \left( en(\xi ) +\partial
\sigma (\xi )/\partial \xi   \right)} \biggr\}\tilde \xi
({\bf{q}}_{\alpha \xi } )\cos \left( {{\bf{q}}_{\alpha \xi }
{\boldsymbol\rho }} \right)\hfill+
\\
+4\pi \sum\limits_{\alpha  = 1}^2 {\frac{{\exp \left( { - (z - \xi
)b(q_{\alpha \sigma } )} \right)}}{{\varepsilon q_{\alpha \sigma }
+ b(q_{\alpha \sigma } )}}\tilde \sigma ({\bf{q}}_{\alpha \sigma }
;\xi )\cos \left( {{\bf{q}}_{\alpha \sigma } {\boldsymbol\rho}}
\right)},\hfill
\\
\tilde \varphi _2 (z,{\boldsymbol\rho}) =  - \sum\limits_{\alpha  =
1}^2 {\frac{{\exp \left( { (z - \xi )q_{\alpha \xi } } \right)}}
{{\varepsilon q_{\alpha \xi }  + b(q_{\alpha \xi } )}}} \biggl\{
{b(q_{\alpha \xi } )}\left(E_1 (\xi )-
\right. \biggr.
\\
\left.- E_2 (\xi )\right)
- \biggl. 
4\pi \left( en(\xi ) +
\partial \sigma(\xi )/\partial \xi 
\right) \biggr\}
\tilde \xi
({\bf{q}}_{\alpha \xi } )\cos \left( {{\bf{q}}_{\alpha \xi }
{\boldsymbol\rho}} \right)+
\\
+4\pi \sum\limits_{\alpha  = 1}^2 {\frac{{\exp \left( { (z - \xi
)q_{\alpha \sigma } } \right)}}{{\varepsilon q_{\alpha \sigma }  +
b(q_{\alpha \sigma } )}}\tilde \sigma ({\bf{q}}_{\alpha \sigma }
;\xi )\cos \left( {{\bf{q}}_{\alpha \sigma } {\boldsymbol\rho}}
\right)}.\hfill
\end{gathered}
\end{equation}
The obtained expressions represent the solution of the potential
distribution problem (so, the charge density distribution, too) in
the area near the dielectric surface with the weak (see
eq.~\eqref{eq77}) spatially periodic inhomogeneities. Let us
emphasize that for the expressions~\eqref{eq78} validity it is no
longer necessary to satisfy the conditions~\eqref{eq71} for all
values of $q$. Its sufficiency is provided by the accomplishment of
the conditions~\eqref{eq71} (or eqs.~\eqref{eq72}, ~\eqref{eq74})
for two-dimensional reciprocal lattice distances $q_{\alpha \sigma
}$, $q_{\alpha \xi }$.

Let us remind that we consider the simplest type of the spatial
periodic inhomogeneities related to the dielectric surface. In the
case of more complicated structure of spatial-periodic homogeneities
it is necessary to use the coefficients of two-dimensional Fourier
series expansion for $ \tilde \xi ({\boldsymbol\rho}) $ and $ \tilde
\sigma ({\boldsymbol\rho};\xi ) $, which describe these
inhomogeneities.

 \section{Conclusion}

Thus, the problem of equilibrium state of the charged particles
above the dielectric surface is solved. Equilibrium distributions
for the charge and electric field induced by these charges in the
system are obtained as in the case of ideally plane dielectric
surface, as in the case of weak spatial inhomogeneities concerned
with the dielectric surface. The weak spatial inhomogeneities caused
as by the inhomogeneities of the surface itself, as by inhomogeneous
charge distributions on it are taken into account. The case of
''wavy'' surface, in particular, the spatially periodic one, is
concerned taking into account the possibility of the surface charge
presence on it. The influence of the external pressing electric
field acting on the system is also taken into consideration. It is
shown that the presence of the gas of volume charges essentially
influences on the value of the electric field potential in the area
near the dielectric surface. Mostly, this fact plays an important
role in the description of the deformation of the liquid dielectric
surface caused by near-surface charges pressure on it. Authors of
the present paper are working on this problem now.

However, in our opinion, the solved problem is useful not only in
two-dimensional Wigner crystallization aspect. The problem is worth
concerning purely with the academic purposes as it can be related to
the number of classical problems of electrodynamics and statistical
physics. Due to this fact, in this paper we do not use the results
of the real experiments on two-dimensional Wigner crystallization
research. The formulations and obtained results in this paper can be
used for the research of the influence of the volume charges near
the liquid helium surface on the spatial inhomogeneous states of
charges, which are adsorbed on the helium surface at the system
parameters close to the experimental ones.

\begin{center}
\textbf{Acknowledgements}
\end{center}
Authors acknowledge financial support from the Consolidated
Foundation of Fundamental Research of Ukraine under grant No.
25.2/102 and thank S.V. Peletminsky for valuable discussions.


\begin{thebibliography}{99}

\bibitem{wigner}
{\it E. Wigner.} On the interaction of electrons in metals. - Phys.
Rev., \textbf{46}, no.11, p. 1002, (1934).

\bibitem{pps}
{\it  A.S. Peletminsky, S.V. Peletminsky, Yu.V.
Slusarenko.} On phase transitions in a Fermi-liquid. II Transition
assotiated with translational symmetry breaking. - Low Temp. Phys.,
\textbf{25}, no.5, p. 303, (1999).

\bibitem{monarkha}
{\it Yu.P. Monarkha and V.B. Shikin.} Low-dimensional electronic systems on a liquid helium surface
(Review). - Sov. J. Low Temp. Phys. \textbf{8}, no. 6, p. 279, (1982). 

\bibitem{mb}
{\it Y. Monarkha and
K. Kono.} Two-dimensional Coulomb liquids and solids. - Berlin:
Springer - Verlag, p. 346, (2003).

\bibitem{cole} {\it M.W. Cole, M.H. Cohen.}
Image-potential-induced surface bands in insulators. - Phys. Rev.
Lett.,  \textbf{23}, no. 21, p. 1238 , (1969).

\bibitem{shykin}
{\it V.B. Shikin.} On helium ions motion near vapour-liquid boundary. - JETP,
\textbf{58}, no. 5, p. 1748, (1970){[in Russian]}.

\bibitem{brown}
{\it T.R. Brown, C.C. Grimes.} Observation of
cyclotron resonance in surface-bound electrons in liquid helium. -
Phys. Rev. Lett., \textbf{29}, no. 18, p. 1233, (1972).

\bibitem{grimes}
{\it C.C. Grimes, T.R. Brown.} Direct spectroscopy observation of electrons in image -
potential state outside liquid helium. - Phys. Rev. Lett.,
\textbf{32}, no.6, p. 280, (1974).

\bibitem{edelman}
{\it V.S. Edelman.} Levitated Electrons. - Sov. Phys. Usp.  \textbf{23} no.4, p.
227,(1980).

\bibitem{tumm}
{\it I.E. Tamm.} Fundamentals of the theory of electricity, Central Books Ltd, 684 pages, (1980).

\bibitem{landau}
{\it  L.D. Landau and E.M. Lifshitz.} \textit{Statistical Physics}, 3rd Ed., Pergamon Pres, Oxford(1981), Nauka, Mascow(1980).

\end{thebibliography}
\end{document}